\documentclass[12pt]{article}
\usepackage[dvips]{graphicx}
\def\be{\begin{equation}}
\def\ee{\end{equation}}
\def\ba{\begin{eqnarray}}
\def\ea{\end{eqnarray}}

\newcommand{\Ea}{\phantom{1}}
\newcommand{\Eb}{\phantom{11}}
\newcommand{\Ec}{\phantom{111}}

\begin{document}
\begin{titlepage}
\thispagestyle{empty}
\vskip0.5cm
\begin{flushright}
MS--TPI--98--5
\end{flushright}
\vskip1.5cm

\begin{center}
{\Large \bf Number of Magic Squares}
\end{center}

\begin{center}
{\Large \bf From Parallel Tempering Monte Carlo}
\end{center}

\vskip1.5cm
\begin{center}
{\large K. Pinn and C. Wieczerkowski}\\
\vskip5mm
Institut f\"ur Theoretische Physik I, Universit\"at M\"unster\\ 
Wilhelm--Klemm--Str.~9, D--48149 M\"unster, Germany \\[3mm]
e--mail: {\sl pinn@uni--muenster.de, wieczer@uni--muenster.de}
\end{center}

\vskip2.5cm
\begin{abstract}
\par\noindent
There are 880 magic squares of size 4 by 4, and 275,305,224 of 
size 5 by 5. It seems very difficult
if not impossible to count exactly the number
of higher order magic squares. 
We propose a method to estimate 
these numbers
by Monte Carlo simulating magic squares at 
finite temperature. One is led to perform low 
temperature simulations of a system with many ground states 
that are separated by energy barriers.  
The Parallel Tempering Monte Carlo method turns
out to be of great help here. 
Our estimate for the number of 6 by 6 magic 
squares is $(0.17745 \pm 0.00016) \times 10^{20}$.
\end{abstract}
\end{titlepage}

\section{Introduction}

Magic squares involve using all the numbers $1, 2, 3, \dots , n^2$ to
fill the squares of an $n \times n$ board so that each row, each
column, and both main diagonals sum up to the same number.  Interesting
information about magic squares is collected at the WEB site of 
M.~Suzuki~\cite{WWW}.

We here address the question of the number of magic squares $N(n)$ of
a given order $n$. In the following, this number should always be
understood as the total number of magic squares divided by 8, thus
considering as equivalent those squares which can be obtained from
each other by the obvious reflection and rotation symmetries. It has
been known since long that there are 880 magic squares of order 4.
$N(5)$ was estimated by L. Candy in his {\em Construction,
Classification and Census of Magic Squares of Order Five}, privately
published in 1938.  Candy arrived at a total of 13,288,952.  The exact
number was determined by Richard Schroeppel in 1973, using a computer
backtracking program, see \cite{Gardner}.  His result 275,305,224
shows that Candy's estimate was low by a wide margin.  It seems very
difficult to exactly determine $N(n)$ for $n > 5$.  However, it is
possible to obtain statistical estimates with good precision. In this
paper we shall describe a Monte Carlo method for this purpose.
As a demonstration, we apply it to the cases $n=4$, 5, and 6.
Our method is, however, by no means restricted to theses special
cases, and could well be used for higher $n$, and also for all
kinds of variants of magic squares, like pan-magic squares, 
squares filled with primes only, or magic cubes.

\section{Magic Squares at Finite Temperature}

We consider magic squares as the zero temperature
configurations of a statistical system with partition function 
\be
Z(\beta) = \sum_{C} \, \exp[-\beta E(C)] \, ,
\ee
where the sum  is over all possibilities to fill the square
with the numbers $n=1,2,\dots,n^2$. $\beta$ is 
proportional to the inverse temperature. We define the 
energy of a configuration by 
\be
E(C)= \sum_{{\rm columns\ } c}   (S_c-M)^2
    + \sum_{{\rm rows\ } r}      (S_r-M)^2
    + \sum_{{\rm diagonals\ } d} (S_d-M)^2 \, ,
\ee
where $S_c$, $S_r$, and $S_d$ are the sums of the columns,
rows, and main diagonals, respectively. $M = n (n^2+1)/2$ 
is called the magic constant.
Obviously $E(C) \geq 0$. Magic squares have zero energy. 
The task of counting the number of magic 
squares $N(n)$ is thus equivalent to counting the number of 
states of minimal energy, i.e. determining the 
zero temperature entropy. 

Monte Carlo methods allow to estimate expectation values of 
functions $A$ of the configurations $C$, 
\be
\langle
A
\rangle 
= \frac{1}{Z} \sum_{C} \, \exp[-\beta E(C)] \, A(C) \, . 
\ee
Now observe that 
\be\label{nnn}
N(n) = \frac18 \lim_{\beta \rightarrow \infty} Z(\beta) \, . 
\ee
$Z(\beta)$ is not an expectation value. However,
since $Z(0)=(n^2)!$, we have  
\be\label{nn1}
N(n) = \frac18 \, Z(0) \, \lim_{\beta \rightarrow \infty}  
\langle \exp[- \beta E(C)] \rangle_{\beta=0} \, ,
\ee
where the subscript indicates that the expectation value 
has to be taken at infinite temperature here. 
Eq.~(\ref{nn1}) is still not practical for 
calculations, since for large $\beta$ the measured quantity 
$\exp[-\beta E(C)]$ fluctuates over many orders of magnitude, 
thus leading to very large statistical errors of the the 
Monte Carlo estimate. 

Let us therefore consider a collection of $\beta$-values 
$0 = \beta_1 < \beta_2 < \dots < \beta_m$. Then 
\be
\frac{Z(\beta_{i+1})}{Z(\beta_i)}
= 
\langle e^{-(\beta_{i+1}-\beta_i)E} \rangle_{\beta_i} \, , 
\ee
so that
\be\label{this}
\frac{Z(\beta)}{Z(0)}= 
\langle e^{-(\beta_2 - \beta_1)\, E} \rangle_{\beta_1} \, 
\langle e^{-(\beta_3 - \beta_2)\, E} \rangle_{\beta_2} \dots
\langle e^{-(\beta_m - \beta_{m-1})\, E} \rangle_{\beta_{m-1}} \, 
\langle e^{-(\beta - \beta_{m})\, E} \rangle_{\beta_{m}}  \, .
\ee 
If the $\beta$-differences in the measured quantities are not
too big, this representations offers a way to compute 
$Z(\beta)$ for large $\beta$ and thus an approximation
for $N(n)$. We remark that $Z(\beta)$ is strictly monotonously 
decreasing. The finite $\beta$-value therefore yields an upper
bound on the number of ground states.

\section{Parallel Tempering Monte Carlo}

A valid Monte Carlo algorithm to estimate any of the expectation values
occurring in eq.~(\ref{this}) can be built using the Metropolis
procedure: Propose to exchange the positions of two entries in the square, 
determine the corresponding energy change $\Delta E$, and implement
the modification of configuration with probability 
\be
p = {\rm min}[1,e^{-\beta \Delta E}] \, . 
\ee
Except for rather small $\beta$, the acceptance rates become prohibitively
small if one just exchanges randomly selected entries. We therefore
restrict the update proposal to the transpositions of 
1 with 2, then 2 with 3, $\dots$, $n^2-1$ with $n^2$. 
Such moves are also useful in Simulated Annealing
procedures designed to search for magic squares with very large $n$, 
by minimizing $E(C)$ or a similar cost function.

For larger $\beta$-values, naive Monte Carlo simulations run into a
problem: The different areas of low energy are separated by high
barriers.  In order to sample all the low energy contributions with
the right weight one has to penetrate and tunnel through these
barries. Consequently, very long simulation times are needed.

The situation can be much improved by using the Parallel Tempering or
Exchange Monte Carlo method, see, e.g. \cite{huku} and further
references cited in \cite{mari}.  It amounts to simulate the joint
ensemble of all the (independent) systems with inverse temperatures
$\beta_i$ in parallel. The partition function of this system is
\be
Z_{\rm joint} = \sum_{C_1} \sum_{C_2} \dots \sum_{C_m}
\exp[ - \beta_1 E(C_1) - \beta_2 E(C_2) - \dots - \beta_m E(C_m)] \, . 
\ee
In addition to updating independently the configurations $C_i$, one 
includes exchanges of configurations, usually of adjacent $\beta$-values.
A proposal of such a change is again accepted using  a Metropolis 
procedure. E.g., configurations $C_i$ and $C_{i+1}$ are exchanged 
with probability  
\be 
p_{i,i+1} = {\rm min}[1,e^{-(\beta_{i+1}-\beta_{i})(E(C_i)-E(C_{i+1}))}] \, .
\ee 
The exchange of configurations over the temperature range strongly
speeds up the Monte Carlo process at the lower temperatures.
Numerical experience shows that,  
in order for the procedure to be efficient, the acceptance rates 
for the configuration exchanges should not be very much smaller
than one half. Furthermore, having too many systems might also 
hamper rapid exchange of information from higher to lower temperatures
and vice versa.

\section{Monte Carlo Results}

We simulated squares with $n=4$, 5 and 6, using always a set of
$m=20$ $\beta$-values. The results are summarized in the tables
\ref{tabL4}, \ref{tabL5}, and \ref{tabL6}.  The largest $\beta$-value
was chosen such that the acceptance rate $\omega_{m}= \omega(\beta_m)$
was around one percent.  The intermediate $\beta$-values were chosen
such that the exchange rates in the tempering cycle are roughly of
order one half or bigger.  Each tempering cycle consisted in
performing one Metropolis updating sweep for each of the 20
configurations and then attempting to exchange each of the adjacent
$\beta_i$-pairs.  We made $ 3.25 \cdot 10^7$ for $n=4$, and $10^8$
such cycles for $n=5$ and $n=6$, respectively. This required
approximately a total of 12 days on a 166 MHz Pentium PC. 
The code was not optimized yet with respect to run-time behaviour.

The $\beta_i$, the acceptance rates $\omega_i$ at $\beta_i$, and the
exchange rates $\omega_{i,i+1}$ are given in columns 2, 3, and 4 of
the tables.  The energy expectation value estimates are given in
column 5.  The last columns of the tables give the ratios of partition
functions $Z(\beta_{i+1})/Z(\beta_i)$, where $Z_i= Z(\beta_i)$.  The
bottom parts present $Z(\beta)/Z(\beta_{20})$ for three extra
$\beta$-values much larger than $\beta_m$. Having three $\beta$'s of
increasing size allows us to check for convergence of the $N(n)$.  The
errors of these estimates were obtained by generating 50 synthetic
data sets of the $Z_{i+1}/Z_{i}$, scattering them around the measured
values according to a Gaussian distribution with variances given by
the error bars of the simulation results.

Both for $n=4$ and $n=5$, our estimates for $N(n)$ stabilize
reasonably with increasing $\beta$ and agree with the exactly known
results. Our estimate for $N(6)$ is $0.17745(16) \cdot 10^{20}$.

\section{Conclusions}

The method proposed provides reliable estimates for the
numbers of magic squares.  Of course, there remain many possibilities to
improve on the simulations (besides going to higher statistics).
E.g., one could play around with $m$, the choices of the $\beta_i$ and
also with the frequency of configuration exchange attemps.

It would be interesting to complement the present approach with, e.g.,
analytical methods. High temperature expansions seem feasable. For
example, the energy at $\beta=0$ can be fairly easily evaluated
exactly, and is given by $\langle E \rangle_0 = n^2 \, (n^4 -1 ) / 6$.
One can convince oneself that the higher moments of the energy at
infinite temperature can all be expressed in closed form, most likely
as polynomials in $n$.

A very short run ($10^6$ cycles) for $n=7$ 
yields $N(7)= 0.3760(52) \cdot 10^{35}$.
It is an interesting question whether there is some simple behaviour
of $N(n)$.  Finally, it could be worthwile to study much larger $n$ to
look out whether the magic squares at finite temperature have also
interesting thermodynamic properties like phase transitions.

\begin{table}
\begin{center}
\begin{tabular}{|r|r|l|l|l|l|}
\hline 
\multicolumn{6}{|c|}{$n=4$ \quad\quad $3.25 \cdot 10^7$ tempering cycles} \\
\hline 
$i\;$ & $\beta_i$ \phantom{xx} &\phantom{x} $\omega_i$  & $\omega_{i,i+1}$ & 
\phantom{xx} $\langle E \rangle_{\beta_i}$ & $Z_{i+1}/Z_i$ \\   
\hline 
 1 &0.0000 & 1.000& 0.868&  680.048(56) & 0.520412(29) \\
 2 &0.0010 & 0.990& 0.922&  626.364(50) & 0.693277(20) \\
 3 &0.0016 & 0.984& 0.861&  594.757(51) & 0.536070(29) \\
 4 &0.0027 & 0.973& 0.800&  539.585(49) & 0.427069(34) \\
 5 &0.0044 & 0.957& 0.713&  463.419(45) & 0.315796(35) \\
 6 &0.0072 & 0.934& 0.631&  365.548(32) & 0.234237(28) \\
 7 &0.0119 & 0.902& 0.582&  262.108(28) & 0.196354(33) \\
 8 &0.0195 & 0.862& 0.552&  176.937(21) & 0.172897(33) \\
 9 &0.0319 & 0.814& 0.529&  114.919(11) & 0.157632(26) \\
10 &0.0523 & 0.754& 0.516& \Ea 72.5736(73) & 0.150015(30) \\
11 &0.0858 & 0.681& 0.508& \Ea 44.9679(50) & 0.148085(27) \\
12 &0.1407 & 0.594& 0.500& \Ea 27.3797(28) & 0.151383(29) \\
13 &0.2308 & 0.491& 0.494& \Ea 16.2867(17) & 0.163141(32) \\
14 &0.3786 & 0.374& 0.495& \Eb  9.3625(11) & 0.188879(35) \\
15 &0.6208 & 0.248& 0.485&  \Eb 5.08151(74)& 0.264513(67) \\
16 &1.0000 & 0.118& 0.704&  \Eb 2.29517(72) & 0.60598(12) \\
17 &1.3000 & 0.057& 0.813&  \Eb 1.15538(60) & 0.78174(11) \\
18 &1.6000 & 0.026& 0.925&  \Eb 0.55536(40) & 0.91600(60) \\
19 &1.8000 & 0.015& 0.951&  \Eb 0.33992(26) & 0.94757(40) \\
20 &2.0000 & 0.009& &        \Eb 0.20941(20) & \\
\hline
\hline  
   & $\beta$ \phantom{xx} &  & &  $Z(\beta)/Z_{20}$ & $N(\beta)/10^3$ \\   
\hline 
   &  5.0  &   & & 0.912727(81) &  0.87968(57)      \\
   &  8.0  &   & & 0.912572(81) &  0.87953(58)      \\
   & 11.0  &   & & 0.912571(81) &  {\bf 0.87953(58)}      \\    
   & exact &   & &              &  {\bf 0.880}    \\
\hline
 \end{tabular}
\parbox[t]{.85\textwidth}
 {
 \caption[tabL4]
 {\label{tabL4}
\small
Monte Carlo results for $n=4$. 
}
}
\end{center}
\end{table}

\begin{table}
\begin{center}
\begin{tabular}{|r|r|l|l|l|l|}
\hline 
\multicolumn{6}{|c|}{$n=5$ \quad \quad $10^8$ tempering cycles} \\
\hline 
$i\;$ & $\beta_i$ \phantom{x} &\phantom{x} $\omega_i$  & $\omega_{i,i+1}$ & 
\phantom{xx} $\langle E \rangle_{\beta_i}$ & $Z_{i+1}/Z_i$ \\   
\hline 
 1 &  0.0000  &  1.000 & 0.561 & 2600.09(17)   & 0.10558(15)\\
 2 &  0.0010  &  0.982 & 0.730 & 1585.362(97)  & 0.29389(21)\\
 3 &  0.0017  &  0.971 & 0.633 & 1188.134(67)  & 0.19578(19)\\
 4 &  0.0029  &  0.955 & 0.551 & \Ea 816.665(45)  & 0.13749(16)\\
 5 &  0.0049  &  0.934 & 0.495 & \Ea 527.305(33)  & 0.10641(13)\\
 6 &  0.0084  &  0.907 & 0.462 & \Ea 327.343(19)  & 0.09006(13)\\
 7 &  0.0142  &  0.873 & 0.442 & \Ea 198.631(11)  & 0.08139(10)\\
 8 &  0.0241  &  0.831 & 0.431 & \Ea 118.9178(64) & 0.07663(10)\\
 9 &  0.0410  &  0.777 & 0.424 & \Eb  70.6331(35) & 0.07403(10)\\
10 &  0.0698  &  0.710 & 0.421 & \Eb  41.7659(21) & 0.0726609(97)\\
11 &  0.1186  &  0.626 & 0.419 & \Eb  24.6289(12) & 0.0719532(95)\\
12 &  0.2016  &  0.525 & 0.418 & \Eb  14.48913(70)& 0.0716943(94)\\
13 &  0.3427  &  0.409 & 0.418 & \Ec   8.50502(44)& 0.0718681(90)\\
14 &  0.5826  &  0.285 & 0.415 & \Ec   4.90719(32)& 0.073181(11) \\
15 &  0.9905  &  0.165 & 0.347 & \Ec   4.90719(32)& 0.093419(26)\\
16 &  1.6838  &  0.058 & 0.810 & \Ec   2.21324(49)& 0.659590(71)\\
17 &  1.9000  &  0.039 & 0.779 & \Ec   1.65545(52)& 0.66774(10)   \\
18 &  2.2000  &  0.021 & 0.881 & \Ec   1.07235(50)& 0.830839(79)  \\
19 &  2.4000  &  0.014 & 0.903 & \Ec   0.79392(47)& 0.871964(76)  \\
20 &  2.6000  &  0.009 &        & \Ec   0.58657(42)&               \\
\hline
\hline  
   & $\beta$ \phantom{xx} &  & &  $Z(\beta)/Z_{20}$ & $N(\beta)/10^9$ \\   
\hline 
   &  5.6000  &   & &  0.66293(25) & 0.27914(19)    \\
   &  8.6000  &   & &  0.65469(26) & 0.27577(19)    \\
   & 11.6000  &   & &  0.65429(26) & {\bf 0.27550(19)}    \\
   & exact    &   & &              & {\bf 0.275305204}    \\
\hline
 \end{tabular}
\parbox[t]{.85\textwidth}
 {
 \caption[tabL5]
 {\label{tabL5}
\small
Monte Carlo results for $n=5$. 
}
}
\end{center}
\end{table}

\begin{table}
\begin{center}
\begin{tabular}{|r|r|l|l|l|l|}
\hline 
\multicolumn{6}{|c|}{$n=6$ \quad\quad $10^8$ tempering cycles} \\
\hline 
$i\;$ & $\beta_i$ \phantom{x} &\phantom{x} $\omega_i$  & $\omega_{i,i+1}$ & 
\phantom{xx} $\langle E \rangle_{\beta_i}$ & $Z_{i+1}/Z_i$ \\   
\hline 
 1 &0.0000 & 1.000 & 0.513&   7768.23(61) &   0.070860(14) \\
 2 &0.0004 & 0.988 & 0.638&   5615.65(36) &   0.168908(18) \\
 3 &0.0008 & 0.979 & 0.502&   4352.01(27) &   0.086142(13) \\
 4 &0.0014 & 0.966 & 0.403&   2976.04(19) &   0.0494970(93) \\     
 5 &0.0027 & 0.948 & 0.345&   1827.89(12) &   0.0342720(72) \\    
 6 &0.0052 & 0.924 & 0.314&   1046.796(62) &  0.027587(58) \\   
 7 &0.0099 & 0.892 & 0.298& \Ea 576.290(33)  &0.0244279(51)\\
 8 &0.0188 & 0.849 & 0.289& \Ea 310.555(17)  &0.0228945(51) \\
 9 &0.0358 & 0.791 & 0.285& \Ea 165.4935(79) &0.0221136(43) \\
10 &0.0679 & 0.713 & 0.283& \Eb  87.6788(38) &0.0216969(42) \\
11 &0.1291 & 0.612 & 0.282& \Eb  46.3095(20) &0.0214833(42) \\
12 &0.2452 & 0.486 & 0.281& \Eb  24.4166(11) &0.0213753(44) \\
13 &0.4660 & 0.340 & 0.282& \Eb  12.86353(59)&0.0213645(47) \\
14 &0.8853 & 0.193 & 0.235& \Ec   6.74920(32)&0.0262622(88) \\
15 &1.6821 & 0.065 & 0.790& \Ec   2.91925(39)&0.570565(54) \\
16 &1.9000 & 0.045 & 0.824& \Ec   2.25025(44)&0.672457(64) \\
17 &2.1000 & 0.031 & 0.844& \Ec   1.73410(48)&0.738535(73) \\
18 &2.3000 & 0.021 & 0.930& \Ec   1.31250(49)&0.884956(43) \\
19 &2.4000 & 0.018 & 0.877& \Ec   1.13603(48)&0.821399(76) \\
20 &2.6000 & 0.012 &       & \Ec   0.84515(46)&             \\
\hline 
\hline  
   & $\beta$ \phantom{xx} &  & &  $Z(\beta)/Z_{20}$ & $N(\beta)/10^{20}$ \\   
\hline 
   &  6.6000  &   & & 0.55845(25)      & 0.17842(16)       \\
   & 10.6000  &   & & 0.55542(25)      & 0.17745(16)       \\
   & 14.6000  &   & & 0.55536(25)      & {\bf 0.17744(16)}       \\
\hline   
 \end{tabular}
\parbox[t]{.85\textwidth}
 {
 \caption[tabL6]
 {\label{tabL6}
\small
Monte Carlo results for $n=6$. 
}
}
\end{center}
\end{table}

\end{document}